\documentclass[twocolumn]{aa}
\usepackage{graphicx}
\usepackage{amsmath,amssymb}
\usepackage{txfonts}

\begin{document}
   \title{Predicted rotation signatures in MHD disc winds\\
     and comparison to DG~Tau observations}

   \subtitle{}

   \author{N. Pesenti\inst{1}, C. Dougados\inst{1}, S. Cabrit\inst{2},
          J. Ferreira\inst{1}, F. Casse\inst{3}, P. Garcia\inst{4},  D. O'Brien\inst{4}}

   \offprints{N. Pesenti}

   \institute{Laboratoire d'Astrophysique UMR 5571, Observatoire de
              Grenoble BP 53, 38041 Grenoble Cedex-9, France\\
              \email{Nicolas.Pesenti@obs.ujf-grenoble.fr} \and LERMA,
              UMR 8112, Observatoire de Paris, 61 avenue de
              l'Observatoire 75014 Paris France \and Institute for
              Plasma Physics Rijnhuizen, P.O. Box 1207, 3430 BE
              Nieuwegein \and Centro de Astrof\'\i sica da Universidade do
              Porto, Rua das Estrelas, 4150-762 Porto, Portugal}

   \date{Received soon ; accepted thereafter}

   \abstract{Motivated by the first detections of rotation signatures
     in the DG~Tau jet (Bacciotti et al. \cite{bacciotti2002}), we
     examine possible biases affecting the relation between detected
     rotation signatures and true azimuthal velocity for self-similar
     MHD disc winds, taking into account projection, convolution as
     well as excitation gradients effects. We find that computed
     velocity shifts are systematically smaller than the true
     underlying rotation curve.  When outer slower streamlines
     dominate the emission, we predict observed shifts increasing with
     transverse distance to the jet axis, opposite to the true
     rotation profile. Determination of the full transverse rotation
     profile thus requires high angular resolution observations ($<
     5$~AU) on an object with dominant inner faster streamlines.
     Comparison of our predictions with HST/STIS observations of
     DG~Tau clearly shows that self-similar, \textit{warm} MHD disc
     wind models with $\lambda = 13$ and an outer radius of the disc
     $\simeq 3$~AU are able to reproduce detected velocity shifts,
     while \textit{cold} disc wind models ($\lambda > 50$) are ruled
     out for the medium-velocity component in the DG~Tau jet.
     
      }
   
   \authorrunning{N. Pesenti et al.}  \titlerunning{Interpreting
     rotation signatures...}  \maketitle
%

\section{Introduction}
\label{intro}

It is now widely accepted that a large scale magnetic field is
responsible for both the acceleration and collimation of jets around
young accreting stars. However, the exact launching zone (stellar
surface, disc truncation radius or wide range of disc radii) remains
subject to debate. So far, studies aimed at constraining proposed
ejection models have concentrated on the jet collimation, poloidal
velocities, and excitation conditions (Dougados et al.
\cite{dougados2000}, Lavalley-Fouquet et al. \cite{lavalley2000},
Bacciotti et al. \cite{bacciotti2000}, Woitas et al.
\cite{woitas2002}).

The detection of rotation signatures in the DG~Tau jet has recently
opened new prospects to constrain MHD ejection models (Bacciotti et
al. \cite{bacciotti2002}). In particular, approximate launching radii
$\simeq 0.3$--$3$~AU were inferred in DG~Tau, indicative of a disc
wind (Anderson et al. \cite{anderson2003}). In this Letter, we
investigate two classes of self-similar disc wind models (Sect. 2) and
analyse in detail the biases introduced in their observed rotation
signatures by projection, beam dilution and emissivity gradients
within the jet (Sect. 3).  Implications for a proper interpretation of
the DG~Tau observations are discussed in Sect. 4.

\section{Disc wind models}

We concentrate our study on two classes of steady, self-similar MHD
keplerian accretion discs driving jets. They are mainly characterized
by 3 nondimensional free parameters (see Cabrit et al.
\cite{cabrit1999}): $\epsilon = h/R_{\mathrm{0}} \simeq 0.1$, where
$h$ is the disc scale height at the disc radius $R_{\mathrm{0}}$,
$\alpha_{\mathrm{m}} \simeq 1$, related to the magnetic diffusivity
parameter, and $\xi \equiv {\rm d}(\log {\dot M}_{\mathrm{acc}})/{\rm
  d}(\log R)$, which controls the mass loading onto field lines. In
these solutions, $\xi$ is related to the magnetic lever arm $\lambda
\simeq (R_{\mathrm{A}}/R_{\mathrm{0}})^2$ ($R_{\mathrm{A}}$ is the
cylindrical radius at the Alfv\'en surface) by the relation $\lambda =
1 + 1/ (2 \xi)$ (Casse \& Ferreira \cite{casse2000a}) and therefore
$\lambda$ does not vary with R$_{\mathrm{0}}$. Ferreira
(\cite{ferreira1997}) obtained \textit{cold} solutions powered by the
Lorentz force only, i.e. where enthalpy is negligible (resulting in
typical $\lambda$ values $\simeq 50$). Detailed comparison with recent
observations showed that these solutions reproduce the collimation
properties of T~Tauri microjets (Dougados et al.
\cite{dougados2000}), but give excessive terminal poloidal velocities
(Garcia et al. \cite{garcia2001b}b).  More recently, Casse \& Ferreira
(\cite{casse2000b}) computed \textit{warm} disc wind solutions where
entropy is injected at the flow base allowing for larger mass loads
($\lambda \simeq 10$) hence lower asymptotic velocities. In the
following, we adopt a \textit{cold} solution with $\lambda = 50$ and a
\textit{warm} solution with $\lambda = 13$. Dimensional scaling
parameters are set as follows: the inner and outer radii of the disc
involved in the ejection process are R$_{\mathrm{i}} = 0.07$~AU
(typical corotation radius for a T~Tauri star) and R$_{\mathrm{e}} =
1$~AU (where molecules are supposed to start to form), the accretion
rate through the disc ${\dot
  M}_{\mathrm{acc}}=10^{-6}$~M$_{\odot}$\,yr$^{-1}$ and the central
stellar mass $M_{\star} = 0.5 M_{\odot}$. The ejection efficiency
($2{\dot M}_{\mathrm{jet}}/{\dot M}_{\mathrm{acc}} = \xi \times \ln
R_{\mathrm{e}}/R_{\mathrm{i}}$) is then $3$\,\% for the \textit{cold}
solution and $10$\,\% for the \textit{warm} one.
  
\begin{figure}
  \begin{center}
    \includegraphics[height=\columnwidth,angle=-90]{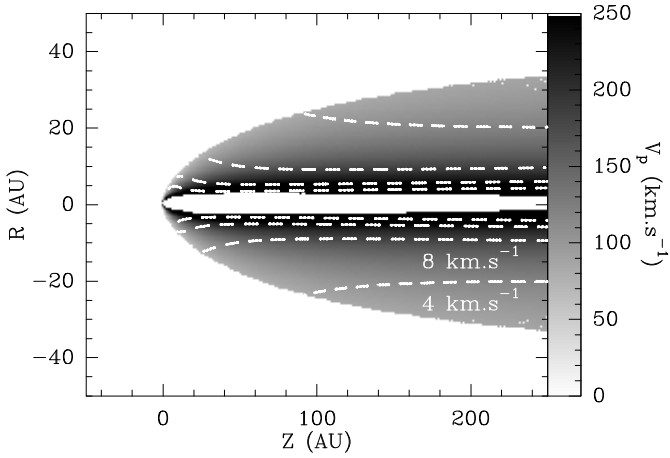}
    \caption{Velocity components ($V_p$,$V_{\phi}$) in the ($R$,$Z$)
      plane for the \textit{warm} solution ($\lambda = 13$).
      \textbf{Dashed lines:} $V_{\phi}$ varying from $4$ to
      $16$~km\,s$^{-1}$ by steps of $4$~km\,s$^{-1}$.
      \textbf{Greyscale:} $V_p$.}
    \label{param}
  \end{center}
\end{figure}

Asymptotic toroidal velocities in MHD disc winds depend mainly on the
magnetic lever arm $\lambda$ and $R_{\mathrm{0}}$.  In Fig.
\ref{param}, we show poloidal and toroidal velocities in the ($R$,$Z$)
plane for the \textit{warm} solution. The \textit{cold} solution shows
a similar map with predicted toroidal and poloidal velocities larger
by a factor $2$. For a given streamline, $V_{\phi}$ rapidly decreases
with distance $Z$ along the jet until the maximum expansion radius is
reached (at $Z/R = 25$ and $20$ for the \textit{cold} and
\textit{warm} solutions respectively).

\begin{figure}
  \begin{center}
    \includegraphics[height=\columnwidth,angle=-90]{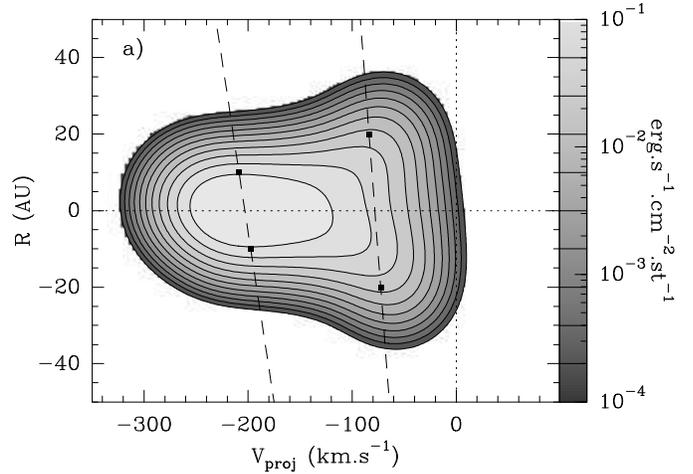}
    \includegraphics[width=\columnwidth,angle=0]{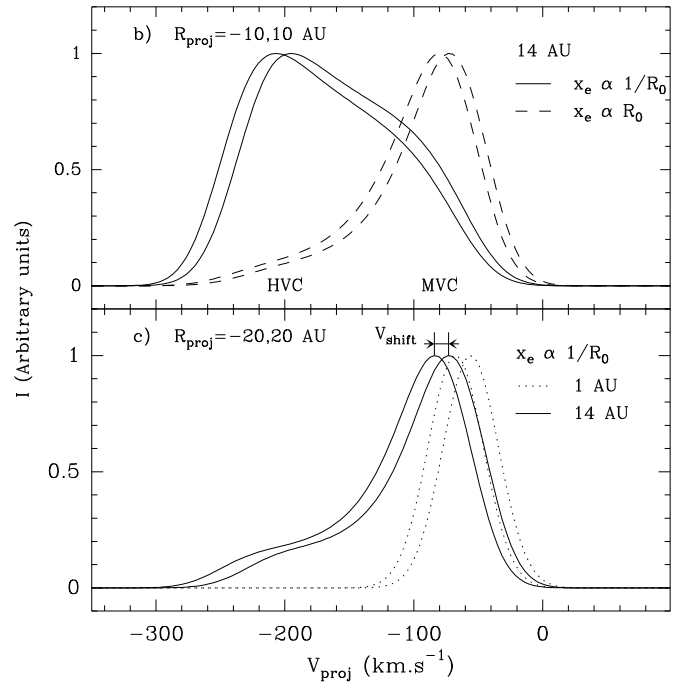}
    \caption{\textbf{a)} Synthetic transverse Position-Velocity (PV)
      diagram for the \textit{warm} solution (integrated over
      $Z_{\mathrm{proj}} = 50$--$60$~AU) in the [O {\sc
        i}]~$\lambda$6300 line, convolved with a $14$~$\mathrm{AU}
      \times 50$~km\,s$^{-1}$ beam. $i = 45^{\circ}$ and
      $x_{\mathrm{e}} = 0.1 \times (0.1 {\rm AU}/R_{\mathrm{0}})$.
      Dashed lines are indicative of peak velocity shifts at $10$ and
      $20$~AU from the jet axis. \textbf{b)} Velocity profiles
      extracted from PV diagrams at $R_{\mathrm{proj}} = \pm 10$~AU,
      convolved with a spatial beam of $14$~AU for $x_{\mathrm{e}} =
      0.1 \times (0.1 {\rm AU}/R_{\mathrm{0}})$ (solid lines) and
      $x_{\mathrm{e}} = 0.1 \times (R_{\mathrm{0}}/1{\rm AU})$ (dashed
      lines), \textbf{c)} and at $R_{\mathrm{proj}}=\pm 20$~AU, with
      spatial beams of $1$ (dotted) and $14$~AU (solid) for
      $x_{\mathrm{e}} = 0.1 \times (0.1 {\rm AU}/R_{\mathrm{0}})$.}
    \label{perp_diag}
  \end{center}
\end{figure}

To calculate predicted rotation signatures for the above models, we
constructed synthetic long-slit spectra perpendicular to the jet axis
(transverse PV diagrams) and convolved them by a two-dimensional
Gaussian beam to simulate the instrumental spatial and spectral
resolutions.  The effect of rotation is to induce a ``tilt'' in
transverse PV diagrams (dashed lines, Fig. \ref{perp_diag}a). We then
get a synthetic velocity shift $V_{\mathrm{shift}}$ by
cross-correlating extracted velocity spectra at symmetric positions
with respect to the jet axis (Fig. \ref{perp_diag}b,c). Our
computations show that $V_{\mathrm{shift}}$ values are only weakly
dependent on the adopted spectral resolution, provided line profiles
are well sampled. We will use here the a velocity resolution of
$50$~km\,s$^{-1}$, which is accessible to current imaging
spectrographs, and smaller than line profile widths.  We adopt a
typical inclination of $i = 45^{\circ}$ with respect to the line of
sight (Pyo et al. \cite{pyo2003}) and consider transverse PV diagrams
at $Z_{\mathrm{proj}} = 50$--$60$~AU.

\begin{figure*}
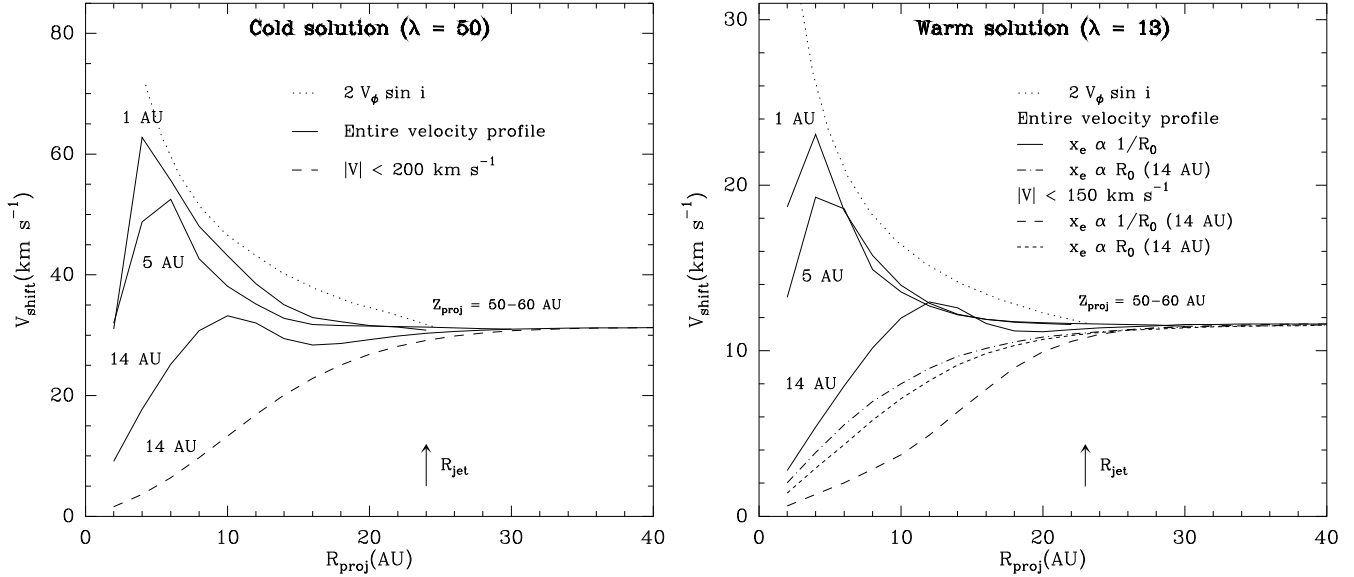

  \begin{center}
    \includegraphics[width=0.9\columnwidth,angle=-90]{Fj061.3a}
    \includegraphics[width=0.9\columnwidth,angle=-90]{Fj061.3b}
    \caption{Velocity shifts predicted by the disc wind models as a
      function of projected transverse radius $R_{\mathrm{proj}}$.
      \textbf{Left:} \textit{cold} solution with the thermal structure
      of Garcia et al. (\cite{garcia2001a}a).  \textbf{Right:}
      \textit{warm} solution with two ionization fraction laws (see
      Sect. 2 for more details). $V_{\mathrm{shift}}$ are obtained by
      cross-correlating [O {\sc i}]~$\lambda$6300 line profiles at
      $\pm R_{\mathrm{proj}}$, extracted from a transverse PV diagram
      at $Z_{\mathrm{proj}} = 50$--$60$~AU. \textbf{Solid and dash-dotted
        lines:} $V_{\mathrm{shift}}$ computed using the entire
      velocity profile, \textbf{short-dashed and long-dashed lines:}
      or restricting to the MVC. \textbf{Dotted lines} show the
      projected underlying true azimuthal profile extracted from the
      corresponding wind solution. $R_{\mathrm{jet}}$ is the outermost
      jet radius (for $R_{\mathrm{e}} = 1$~AU).}
    \label{Vshift}
  \end{center}
\end{figure*}

We will use the [O {\sc i}]~$\lambda$6300 line, the effect of a tracer
with lower critical density being discussed later. We explore spatial
resolution effects by varying the beam size (FWHM) between 1~AU and
14~AU. To calculate line emissivities in the \textit{cold} solution,
we use the thermal and ionization structure computed \textit{a
  posteriori} by Garcia et al. (\cite{garcia2001a}a) with ambipolar
diffusion heating. The resulting $T_{\mathrm{e}}$ is roughly constant
($T_{\mathrm{e}} \sim 10^4$~K) and the ionization stratification is
close to a $x_{\mathrm{e}} \propto 1/R_{\mathrm{0}}$ law. This gives
line profiles with a strong high velocity component (HVC), originating
from the inner faster streamlines (Garcia et al. \cite{garcia2001b}b).
In the case of the \textit{warm} solution, where the full thermal
solution is not yet available, we keep $T_{\mathrm{e}} = 10^4$~K and
we explore two extreme ionization fraction laws: one similar to the
\textit{cold} solution, $x_{\mathrm{e}} = 0.1 \times (0.1 {\rm
  AU}/R_{\mathrm{0}})$, where the HVC dominates (solid curves in Fig.
\ref{perp_diag}b), and one with $x_{\mathrm{e}} = 0.1 \times
(R_{\mathrm{0}}/1{\rm AU})$, where emission from outer slower
streamlines is enhanced, and the profile is dominated by a medium
velocity component (MVC, dashed curves, Fig.  \ref{perp_diag}b). The
latter case reproduces better observed line profiles in the DG~Tau jet
(see Bacciotti et al. \cite{bacciotti2002}), and allows us to
investigate the effect of various emissivity gradients in the jet on
the observed $V_{\mathrm{shift}}$.

\section{Detailed analysis of disc wind predictions}

We first discuss $V_{\mathrm{shift}}$ values obtained by
cross-correlation over the entire velocity profile, plotted in Fig.
\ref{Vshift}. We also plot for comparison as a dotted curve the
projected azimuthal velocity profile
($V_{\mathrm{shift}}^{\mathrm{theo}}(R) = 2 \times V_{\phi}(R) \times
\sin{i}$) extracted from the corresponding wind solution at the same
$Z_{\mathrm{proj}}$.  We find that measured velocity shifts always
underestimate the true rotation profile, especially at small
transverse distances from the jet axis. This underestimate comes from
the integration of the flow along the line of sight, combined with
projection effects: At a given radius $R_{\mathrm{proj}}$ from the jet axis,
the line of sight intersects a range of flow surfaces with true
cylindrical radii $R>R_{\mathrm{proj}}$.  Due to the keplerian law, they rotate
more slowly than $V_{\phi}(R_{\mathrm{proj}})$. In addition, these
outer surfaces are not tangent to the line of sight and their
projected rotation speed is $\sin\,i \times V_{\phi}(R) \times
\sin\phi$ with $\sin\phi = R_{\mathrm{proj}}/R$ introducing an
additional reduction factor. However, the exact amount by which the
rotation profile is underestimated by these projection effects
critically depends on excitation gradients and beam dilution, as we
now discuss.

We first consider the case where inner streamlines (HVC) dominate line
profiles, which is illustrated by the \textit{cold} solution and the
\textit{warm} solution with $x_{\mathrm{e}} \propto 1/R_{\mathrm{0}}$
(solid curves, Fig.  \ref{Vshift}). A given beam tends to decrease
$V_{\mathrm{shift}}$ at projected radii below the beam diameter (see
14 AU beam, Fig. \ref{Vshift}) because of cancellation effects between
opposite sides of the jet. When $R_{\mathrm{proj}}$ varies from the
beam diameter to the jet outer radius $R_{\mathrm{jet}}$, each profile
is dominated by the surface at $R = R_{\mathrm{proj}}$ and the
discrepancy from the theoretical value is small. Only at radii smaller
than the central hole radius $R_{\mathrm{hole}}$ ($4$~AU for both
solutions), $V_{\mathrm{shift}}$ tend to zero as
$R_{\mathrm{proj}}/R_{\mathrm{hole}}$, due to the above projection
effect.

We examine now the case where outer streamlines (MVC) dominate line
profiles, which is illustrated by the \textit{warm} solution with
$x_{\mathrm{e}} \propto R_{\mathrm{0}}$ (dash-dotted curves, Fig.
\ref{Vshift}). Line profiles at $R_{\mathrm{proj}}$ are now strongly
contaminated by surfaces with true radii $R > R_{\mathrm{proj}}$, so
that $V_{\mathrm{shift}}$ lies significantly below
$V_{\mathrm{shift}}^{\mathrm{theo}}$ (a factor $2$ at $10$~AU),
regardless of beam size. Only $V_{\mathrm{shift}}$ measured at
projected distances on the order of $R_{\mathrm{jet}}$ fit correctly
$V_{\mathrm{shift}}^{\mathrm{theo}}$, the projection factor
$R_{\mathrm{proj}}/R$ being then $\sim 1$.

Finally, when cross correlation is made on the MVC only (long-dashed
and short-dashed curves, Fig. \ref{Vshift}), $V_{\mathrm{shift}}$ is
even lower and varies almost as $R_{\mathrm{proj}}/R_{\mathrm{jet}}$.
This can be understood by considering that cross-correlation on the
MVC is sensitive only to a narrow range of outer streamlines at $R
\sim R_{\mathrm{jet}}$. Beam smearing and excitation gradients have
little effect because this region is far from the jet axis and has
roughly homogeneous excitation conditions.

We have verified that the above results remain valid over a broad
region of the parameter space. Variation of $M_{\star}$,
$R_{\mathrm{i}}$ and $R_{\mathrm{e}}$ only introduces scaling factors
($V \propto \sqrt{M_{\star}/R_{\mathrm{0}}}$), $R_{\mathrm{hole}}$ and
$R_{\mathrm{jet}}$ being about proportional to $R_{\mathrm{i}}$ and
$R_{\mathrm{e}}$ respectively.  $V_{\mathrm{shift}}$ obtained from a
solution with $\lambda = 8$ have the same biases as those described
above, suggesting that our results also do not depend critically on
the detailed disc wind model. The only other parameter of significant
influence on the predicted $V_{\mathrm{shift}}$ is the electronic
density to critical density ratio ($N_{\mathrm{e}}/N_{\mathrm{cr}} =
N_{\mathrm{H}} x_{\mathrm{e}} / N_{\mathrm{cr}}$ with
$N_{\mathrm{H}}$, the total density) for a given line. If
$N_{\mathrm{e}}/N_{\mathrm{cr}} > 1$ in the central parts of the jet,
either because the accretion rate is increased (${\dot
  M}_{\mathrm{acc}} \propto N_{\mathrm{H}}$) or a tracer with low
$N_{\mathrm{cr}}$ (such as [S~{\sc ii}]~$\lambda$6731) is used, the
emissivity saturates and $V_{\mathrm{shift}}$ becomes significantly
lower than $V_{\mathrm{shift}}^{\mathrm{theo}}$ (though not as much as
for MVC-dominated profiles). A ratio $N_{\mathrm{e}}/N_{\mathrm{cr}} <
1$ in the inner part of the jet, and thus a tracer with high
$N_{\mathrm{cr}}$, is needed to retrieve the shape of
$V_{\mathrm{shift}}$ curves shown in Fig.  \ref{Vshift} in the
HVC-dominated case for high accretion rates.

\section{Comparison with DG~Tau observations}

\begin{figure}
  \begin{center}
    \includegraphics[width=\columnwidth]{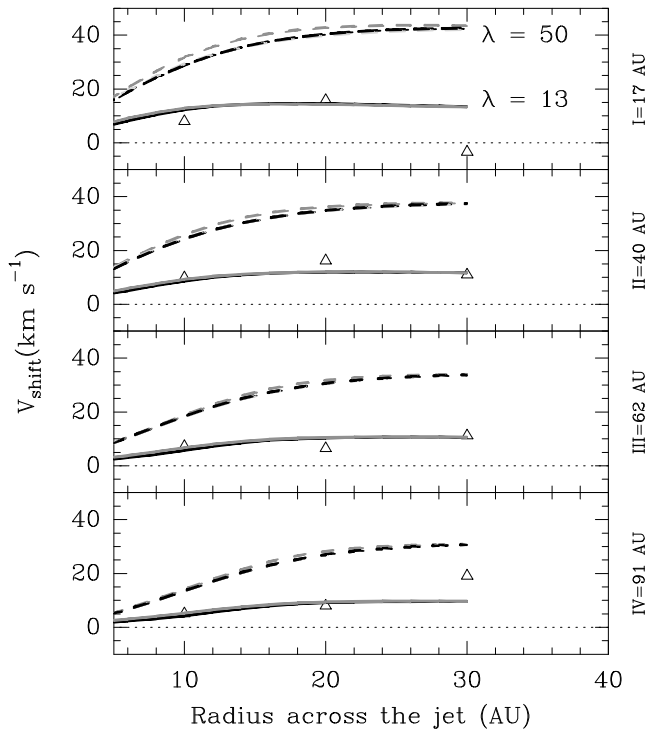}
    \caption{Velocity shifts as a function
      of radius at four deprojected distances from the star, 17 to
      91~AU, obtained by cross-correlation over the MVC.
      \textbf{Symbols:} DG~Tau observations (Bacciotti et al.  2002).
      \textbf{Curves:} Predictions for the \textit{cold} (dashed; $R_{\mathrm{e}}
      =1$~AU) and \textit{warm} (solid; $R_{\mathrm{e}} = 3$~AU) disc wind
      solutions in [S {\sc ii}]~$\lambda$6731 (light grey), [O {\sc
        i}]~$\lambda$6300 (dark grey) and [Fe {\sc ii}]~1.644~$\mu$m
      (black), computed with the estimated inclination angle of DG~Tau
      jet (45$^{\circ}$; Pyo et al. \cite{pyo2003}) and with the beam
      size of HST/STIS (0.1$''$\,$\times$\,50\,km\,s$^{-1}$).}
    \label{bacciotti}
  \end{center}
\end{figure}

We compare here our predictions with observations of DG~Tau carried
out by Bacciotti et al. (\cite{bacciotti2002}) with HST/STIS.
$V_{\mathrm{shift}}$ were obtained from the MVC by cross-correlation
techniques similar to the one described in Sect. 2. In Fig.
\ref{bacciotti}, we plot observed $V_{\mathrm{shift}}$ (symbols) and
predictions for both disc wind models with an equivalent beam size
($0.1^{\prime\prime} \times 50$~km\,s$^{-1}$) and at the same
distances (deprojected) from the star.  Predicted velocity shifts are
plotted in several spectral lines, and are seen to differ by less than
$10$\,\%.
 
We first note that both the \textit{cold} and \textit{warm} solutions
reproduce well the trend of increased measured $V_{\mathrm{shift}}$
with transverse radius in DG~Tau. Based on the above study, we
interpret this trend as due to projection effects only. Thus,
measurements at $R_{\mathrm{proj}} = 10,20$~AU $< R_{\mathrm{jet}}(Z)$
do not yield direct measures of $V_{\phi}$, but only lower limits.
Only data points at $R_{\mathrm{proj}} = 30$~AU, close to the outer
jet radius, would be expected to give true $V_{\phi}$.

Second, we note that $V_{\mathrm{shift}}$ deduced from the
\textit{cold} disc wind model $\simeq 30$--$40$~km\,s$^{-1}$ are about
by a factor $3$--$6$ too high compared to observed velocity shifts of
$6$--$15$~km\,s$^{-1}$. Steady, self-similar \textit{cold} solutions
with lower toroidal velocities (lower $\lambda$) exist, but terminate
too close to the star ($Z/R_{\mathrm{0}} \le 10$; Ferreira
\cite{ferreira1997}).  This class of solutions is therefore definitely
excluded for the MVC in the DG~Tau jet. On the other hand, the
\textit{warm} disc wind solution with $\lambda = 13$ fit very well
observed $V_{\mathrm{shift}}$ in the DG~Tau jet within instrumental
uncertainties, except for the $30$~AU datapoints in region I and IV. Far
from the jet axis, region I suffer from a poor signal-to-noise ratio,
while region IV may be contaminated by non-axisymmetric structures in
the external medium and/or bowshocks, similar to those at larger scale
(Lavalley-Fouquet et al.  \cite{lavalley2000}).

Although the difference in $V_{\mathrm{shift}}$ between
$R_{\mathrm{e}} = 1$ and $3$~AU is less than $25$\,\%, we note that
the latter $R_{\mathrm{e}}$ value is essential for reproducing
observed centroid velocities $\simeq 50$~km\,s$^{-1}$ of the MVC in
DG~Tau ($\simeq 80$~km\,s$^{-1}$ would be found for $R_{\mathrm{e}} =
1$~AU; see Fig. \ref{perp_diag}b). This value is consistent with
results of the diagnostic method of disc wind launching radii proposed
by Anderson et al. (\cite{anderson2003}). However, launching radii
$R_{\mathrm{0}} = 0.3$--$1$~AU deduced with this method from the
underestimated azimuthal velocities at $R_{\mathrm{proj}} = 10,20$~AU,
are only lower limits.  Observations with a spatial beam $< 5$~AU and
high-density tracers for ionized gas, favouring HVC-dominated line
profiles, would allow to derive the full transverse velocity profile
($\propto 1/\sqrt{R}$), and also more constraints on $\lambda$,
$R_{\mathrm{i}}$ and $R_{\mathrm{e}}$.  Such a resolution is currently
achievable in the UV domain with HST/STIS and in the near-IR domain
with the new generation of adaptive optics system operating on 8m
class telescopes such as NACO on the VLT.

\begin{acknowledgements}
We would like to thank the referee, F. Bacciotti, for her very helpful comments.
\end{acknowledgements}

\end{document}